\documentclass{article}
\usepackage{spconf,amsmath,graphicx}

\usepackage{enumitem}
\setlist{nosep, leftmargin=14pt}

\usepackage{mwe} % to get dummy images

\usepackage{booktabs}      % for \toprule, \midrule, \bottomrule
\usepackage{array}         % for better column alignment and spacing control
\usepackage{multirow}      
\usepackage{siunitx}       % for alignment of numbers using S columns
\usepackage{subcaption}
\usepackage{amsfonts}       % or \usepackage{amssymb}
\usepackage{placeins}   % for \FloatBarrier
\usepackage{stfloats}   % better control of two-column floats
\usepackage{float}   % enables [H] for non-floating placement
\usepackage{cuted}   % allows full-width, non-floating tables/figures

\title{Brain--MGF: Multimodal Graph Fusion Network for EEG--fMRI Brain Connectivity Analysis under Psilocybin}

\name{\begin{tabular}{c}
Sin-Yee Yap$^{1}$, Fuad Noman$^{1}$,  Junn Yong Loo$^{1}$,  Devon Stoliker$^{3}$, Moein Khajehnejad$^{3}$,\\
Rapha\"el C.-W. Phan$^{1}$, David L. Dowe$^{2}$, Adeel Razi$^{3,\star}$, and Chee-Ming Ting$^{1,\star}$
\end{tabular}}

\address{$^{1}$ School of Information Technology, Monash University Malaysia, Malaysia \\
         $^{2}$ Department of Data Science and AI, Faculty of Information Technology, Monash University, Australia \\
         $^{3}$ Turner Institute for Brain and Mental Health, School of Psychological Sciences, Monash University, Australia}

\begin{document}
\ninept
\maketitle

% Corresponding authors
\begingroup
\renewcommand\thefootnote{*}
\footnotetext{Corresponding authors: Adeel Razi (\texttt{adeel.razi@monash.edu}) and Chee-Ming Ting (\texttt{ting.cheeming@monash.edu})}
\endgroup

\begin{abstract}
Psychedelics, such as psilocybin, reorganise large-scale brain connectivity, yet how these changes are reflected across electrophysiological (electroencephalogram, EEG) and haemodynamic (functional magnetic resonance imaging, fMRI) networks remains unclear. We present \textbf{Brain--MGF}, a multimodal graph fusion network for joint EEG–fMRI connectivity analysis. For each modality, we construct graphs with partial-correlation edges and Pearson-profile node features, and learn subject-level embeddings via graph convolution. An adaptive softmax gate then fuses modalities with sample-specific weights to capture context-dependent contributions. Using the world's largest single-site psilocybin dataset, \textit{PsiConnect}, Brain--MGF distinguishes psilocybin condition from no-psilocybin condition in meditation and rest. Fusion improves over unimodal and non-adaptive variants, achieving \textbf{74.0\%} accuracy and \textbf{76.5\%}
F1 score on meditation, and \textbf{76.0\%} accuracy with \textbf{85.8\%} ROC–AUC on rest. UMAP visualisations reveal clearer class separation for fused embeddings. These results indicate that adaptive graph fusion effectively integrates complementary EEG-fMRI information, providing an interpretable framework for characterising psilocybin-induced alterations in large-scale neural organisation.
\end{abstract}
\begin{keywords}
Multimodal Graph Fusion, EEG–fMRI, Brain Connectivity, Psychedelics, Psilocybin
\end{keywords}

\section{Introduction}
\label{sec:intro}

% Psilocybin?
Psilocybin, the psychoactive compound in \textit{Psychedelic} mushrooms, serves as a tool to perturb perception for investigating altered states of consciousness. It induces transient changes in perception and cognition accompanied by large-scale reorganisation of brain networks. Studies report reduced default-mode network connectivity and increased cross-network communication \cite{stoliker2025psychedelics, gattuso2023default}, reflecting greater global integration. However, how these network-level alterations manifest across physiological (temporal) scales remains unclear, despite their relevance to therapeutic mechanisms linking reduced modularity with improved mental health \cite{kuburi2022neuroimaging,daws2022increased}. In this paper, we
investigate whether psilocybin-induced brain states can be distinguished from no-psilocybin states in the same participants across distinct experimental stimuli using graph-based multimodal connectivity representations, providing a computational approach to quantify how psychedelics reorganise large-scale neural integration. 

% GNN
Analysing these network-level effects requires models that capture the brain’s graph-like organisation and distributed connectivity. Recent advances in network neuroscience emphasise the importance of modelling the brain as a complex graph of interacting regions rather than a collection of isolated areas \cite{bassett2017network}. Graph neural networks (GNNs) have emerged as powerful tools for analysing such structured data, offering the ability to learn representations that capture both local and global connectivity patterns \cite{bessadok2022graph}. In brain imaging, GNNs have demonstrated strong performance in tasks such as disease diagnosis \cite{li2021braingnn,noman2022graph,noman2025adaptive}, cognitive state decoding \cite{zhang2021functional,ye2023explainable}, and behavioural prediction \cite{wu2021connectome,wen2024multi}, as they can encode topological dependencies that traditional machine learning models overlook. 

% Multimodal
Building upon this, multimodal neuroimaging enables the investigation of brain connectivity across complementary physiological domains. Functional magnetic resonance imaging (fMRI) captures large-scale haemodynamic coupling, at a temporal scale of seconds, that reflects distributed integration among cortical and subcortical regions owing to its higher spatial resolution, whereas electroencephalography (EEG) measures, at the millisecond-scale, electrophysiological coherence that indexes fast-scale neuronal synchrony, however without the spatial resolution to resolve signals from sub-cortical regions \cite{shibasaki2008human}. Integrating these complementary connectivity profiles might therefore yield a more comprehensive characterisation of psilocybin’s network-level effects by linking haemodynamic and electrophysiological coupling patterns.
%However, EEG and fMRI differ in spatial coverage, signal characteristics, and the scale of connectivity they capture, posing methodological challenges for multimodal integration.

% Challenges?
While such multimodal data promise richer insights, computational strategies for combining them are still evolving and require further methodological refinement. Conventional approaches typically concatenate features or assign fixed weights to modalities \cite{baltruvsaitis2018multimodal}, neglecting the
%heterogeneity of neural responses across individuals and experimental contexts (e.g., different task stimuli and pharmacological states) 
inter-individual variability and differences in task stimuli or pharmacological state.
%\cite{farahani2019application}.
Such fixed fusion strategies are limited in their ability to capture the dynamic, subject-specific contributions of each modality. Existing multimodal GNNs often perform late fusion without adaptive weighting \cite{mohammadi2024graph},  failing to account for the varying relevance of each modality across cognitive states or participants \cite{sui2023data}. These considerations suggest that adaptively learning modality contributions may offer a more suitable and context-responsive approach than fixed fusion schemes.
%An adaptive fusion approach is therefore essential to learn modality contributions in a data-driven and context-sensitive manner.

% Contributions
To address this gap, we introduce the \textbf{Brain--MGF (Multimodal Graph Fusion Network)}, an adaptive GNN framework for the joint analysis of EEG and fMRI connectivity. Brain--MGF employs modality-specific graph encoders to extract latent representations and a gating mechanism that dynamically modulates the contribution of each modality during fusion. This architecture enables flexible integration of complementary neural information across physiological scales, facilitating a more precise characterisation of brain dynamics.
%, for example, under psychedelic modulation.
Our main contributions are as follows:
\begin{itemize}[leftmargin=12pt]
    \item \textbf{Adaptive multimodal fusion:} a gating-based graph model that learns subject-specific modality weights, enabling context-dependent integration of EEG and fMRI signals.
    \item \textbf{Graph construction with complementary features:} edges derived from partial correlations and node features from Pearson profiles with a pseudo-identity projection, jointly representing both direct dependencies and global co-activation patterns.
    \item \textbf{Empirical validation on the \textit{PsiConnect dataset}:} evaluation %on the meditation condition
    in participants imaged during guided meditation and rest demonstrates that Brain--MGF outperforms unimodal and non-adaptive variants, producing more separable latent representations that reflect enhanced discriminability between psilocybin and no-psilocybin brain states.
\end{itemize}

\FloatBarrier

\section{Methods}
\label{sec:methods}

% Figure: Brain-MGF Architecture
\begin{figure*}[t]
\centering
\includegraphics[width=0.95\textwidth]{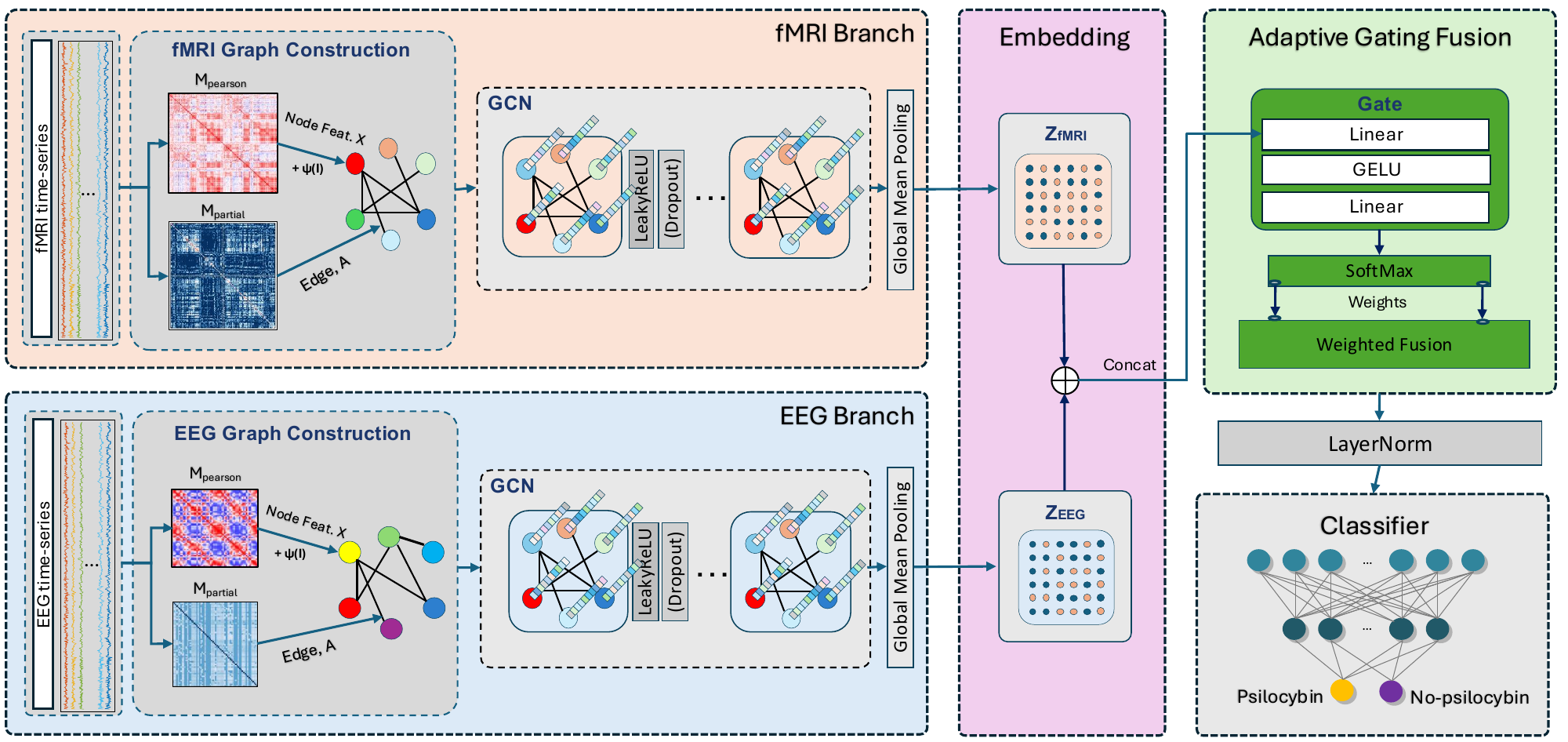}
\caption{\textbf{Brain--MGF architecture.} 
Each modality (fMRI and EEG) constructs a connectivity graph from Pearson and partial correlations, combined with a pseudo-identity projection to form node features $\mathbf{X}$ and edges $\mathbf{A}$. 
Three GraphConv blocks with LeakyReLU and dropout encode node representations, followed by global mean pooling to obtain subject-level embeddings $\mathbf{z}_{\mathrm{fMRI}}$ and $\mathbf{z}_{\mathrm{EEG}}$. 
An adaptive gating MLP computes softmax weights $\alpha_{\mathrm{fMRI}}, \alpha_{\mathrm{EEG}}\!\in\!(0,1)$ to fuse modalities: 
$\mathbf{z}_{\mathrm{fus}}=\alpha_{\mathrm{fMRI}}\mathbf{z}_{\mathrm{fMRI}}+\alpha_{\mathrm{EEG}}\mathbf{z}_{\mathrm{EEG}}$. 
The fused embedding is normalised with LayerNorm and passed through an MLP classifier with sigmoid output, optimised using binary cross-entropy loss.}
\label{fig:brain_mgf}
\vspace{-2mm}
\end{figure*}

\subsection{Dataset and Preprocessing}
We used the \textit{PsiConnect} dataset \cite{novelli2025psiconnect,stoliker2025psychedelics}, which contains EEG and fMRI recorded under psilocybin and no-psilocybin across naturalistic conditions (\textit{rest}, \textit{meditation}, \textit{music}, and \textit{movie}). In this paper, we focused on rest and guided meditation because these contexts exhibited lower inter-subject variability and stronger cross-modal alignment between EEG and fMRI connectivity patterns, compared to the more stimulus-driven music and movie conditions.
The study involved 65 healthy adults (aged 18–55). After quality control, fMRI data were available for 64 participants in the no-psilocybin session and 61 in the psilocybin session, while EEG data were available for 64 and 59 participants, respectively.
%After quality control, fMRI data were available for 64 participants at no-psilocybin and 61 at psilocybin, while EEG data were available for 64 at no-psilocybin and 59 at psilocybin.
All data followed the \textit{PsiConnect} preprocessing pipeline using \texttt{fMRIPrep} \cite{esteban2019fmriprep} and the \texttt{RELAX} EEG pipeline \cite{bailey2023introducing}. For fMRI, preprocessing included slice-timing and motion correction, spatial normalisation to  MNI (Montreal Neurological Institute) space, and temporal filtering (0.01--0.1~Hz). Regional
BOLD
(Blood Oxygenation Level–Dependent)
signals were extracted using the Schaefer 200-parcel atlas (7-network version). For EEG, artefacts were removed using multi-channel Wiener filtering and wavelet-enhanced ICA, followed by robust re-referencing. Each cleaned recording was arranged as a time~$\times$~channel array (64 channels, 500~Hz).

\subsection{Graph Construction}
Each subject and modality was represented as a weighted undirected graph 
$G = (V, E, \mathbf{A})$,
% where
with
$|V|=200$ for
% fMRI
fMRI,
% and
$|V|=64$ for EEG. 

For fMRI, regional BOLD time series 
$\mathbf{X}\in\mathbb{R}^{T\times200}$ were extracted using the Schaefer 200-parcel atlas \cite{schaefer2018local}. 
Functional connectivity was estimated using both Pearson and partial correlations between regions. 
The Pearson correlation between regions $i$ and $j$ is defined as
\begin{equation}
% M_{\mathrm{pearson},ij} =
% \frac{\mathrm{cov}(x_i,x_j)}{\sigma(x_i)\sigma(x_j)},
M_{\mathrm{pearson},ij} =
{\mathrm{cov}(x_i,x_j)} / {(\sigma(x_i)\sigma(x_j))},
\label{eq:pearson}
\end{equation}
where $x_i$ and $x_j$ denote the BOLD time series of regions of interest (ROIs) $i$ and $j$, and $\mathbf{\Sigma}$ denotes the covariance matrix across all ROIs. 
Partial correlations were computed from the inverse covariance matrix 
$\mathbf{\Theta}=\mathbf{\Sigma}^{-1}$ as
\begin{equation}
% M_{\mathrm{partial},ij} =
% -\frac{\Theta_{ij}}{\sqrt{\Theta_{ii}\Theta_{jj}}},
M_{\mathrm{partial},ij} =
- {\Theta_{ij}} / (\sqrt{\Theta_{ii} \Theta_{jj}}),
\label{eq:partial}
\end{equation}
which capture direct functional dependencies after removing indirect effects mediated by other regions. 
Each resulting $200\times200$ matrix was symmetrised with a unit diagonal and sparsified by retaining the top 40\% of the strongest connections to preserve salient edges.
%via top-$k$ thresholding ($k=40$) to retain the most salient edges.

For EEG, functional connectivity was derived from the time~$\times$~channel matrix 
$\mathbf{S}\in\mathbb{R}^{T\times64}$ by computing Pearson and partial correlations across channels, 
using the same definitions as in Eqs.~\eqref{eq:pearson}–\eqref{eq:partial}. 
Each resulting $64\times64$ matrix was symmetrised and sparsified in the same way, retaining top 40\% of the strongest connections.
%using top-$k$ thresholding ($k=40$).

%\subsection{Benchmark Comparison}
% Table: Benchmark Comparison
\begin{table*}[b]
\centering
\setlength{\tabcolsep}{3pt}
\renewcommand{\arraystretch}{1.05}
\caption{Benchmark comparison of unimodal (fMRI) and multimodal (EEG+fMRI) models on the meditation task. 
Brain-MGF denotes the proposed Adaptive Graph Fusion Network and its ablated variants. 
Results are reported as mean~$\pm$~SD over 5 folds (BCE loss).}
\label{tab:benchmark_meditation}
\footnotesize
\begin{tabular}{lcccccccc}
\toprule
\textbf{Model} & \textbf{Modality} & \textbf{Accuracy} & \textbf{Precision} & \textbf{Sensitivity} & \textbf{Specificity} & \textbf{F1 Score} & \textbf{ROC-AUC} & \textbf{BCE Loss} \\
\midrule
\textbf{GCN \cite{kipf2016semi}} & fMRI & 69.71$\pm$5.73 & 68.92$\pm$4.19 & 79.23$\pm$11.44 & 58.79$\pm$12.65 & 73.20$\pm$5.88 & 76.45$\pm$9.67 & 0.6358$\pm$0.2090 \\
BrainGNN \cite{li2021braingnn} & fMRI & 61.27$\pm$7.15 & 66.22$\pm$9.50 & 56.92$\pm$5.65 & 65.91$\pm$13.64 & 60.96$\pm$6.34 & 67.12$\pm$7.93 & 0.6514$\pm$0.0845 \\
BrainNetCNN \cite{kawahara2017brainnetcnn} & fMRI & 62.93$\pm$8.14 & 64.46$\pm$12.77 & 82.69$\pm$17.29 & 41.21$\pm$29.23 & 69.89$\pm$5.73 & 69.58$\pm$6.64 & 0.8479$\pm$0.4399 \\
\cmidrule(lr{0.25em}){1-9}
\textbf{Late-Fusion (MLP)} & EEG+fMRI & 69.75$\pm$3.09 & 68.63$\pm$4.11 & 80.64$\pm$8.89 & 56.97$\pm$13.74 & 73.69$\pm$3.02 & 74.46$\pm$2.65 & 0.6567$\pm$0.1318 \\
\textbf{Brain-MGF (w/o Gate)} & EEG+fMRI & 69.67$\pm$6.54 & 70.50$\pm$7.30 & 74.36$\pm$12.59 & 64.55$\pm$11.64 & 71.71$\pm$8.10 & \textbf{77.13}$\pm$9.21 & 0.6459$\pm$0.1675 \\
\textbf{Brain-MGF (Proposed)} & EEG+fMRI & \textbf{74.02}$\pm$6.99 & \textbf{74.05}$\pm$9.00 & \textbf{80.90}$\pm$12.93 & \textbf{66.21}$\pm$17.22 & \textbf{76.48}$\pm$6.95 & 75.72$\pm$11.29 & \textbf{0.5800}$\pm$0.1075 \\
\bottomrule
\end{tabular}
\end{table*}
% comment: DLD comments to again ask what    Loss    is supposed to be

In the Brain--MGF model, partial-correlation matrices $M_{\mathrm{partial}}$ define the graph structure,
while Pearson matrices $M_{\mathrm{pearson}}$ provide node-wise connectivity profiles as features. 
For each modality, the node feature matrix is given by
\begin{equation}
\mathbf{X} = M_{\mathrm{pearson}} + \psi(\mathbf{I})
\end{equation}
where $\psi(\cdot)$ is a learnable linear projection applied to the identity matrix $\mathbf{I}$ to introduce a 
pseudo-identity prior. 
In implementation, this corresponds to
\texttt{node\_feat = Mpearson + fc\_p(pseudo)}, 
where \texttt{fc\_p} denotes the learnable projection layer.
Thus, the GNN operates on node features $\mathbf{X}$ and sparsified partial-correlation adjacency 
%$\tilde{\mathbf{A}}=\mathcal{T}_k(M_{\mathrm{partial}})$, combining global co-activation (Pearson) and direct dependencies (partial correlation) within a unified graph representation.
$\tilde{\mathbf{A}}=\mathcal{T}_{40\%}(M_{\mathrm{partial}})$, 
where $\mathcal{T}_{40\%}(\cdot)$ denotes a thresholding operator that retains the top 40\% of edges 
(by absolute partial-correlation strength) while setting the rest to zero. 
This combines global co-activation (Pearson) and direct dependencies (partial correlation) within a unified graph representation.

\subsection{Proposed Brain--MGF Framework}

The proposed Brain--MGF integrates modality-specific graph encoders with an adaptive fusion gate for joint EEG--fMRI analysis. The model architecture is shown in Fig.~\ref{fig:brain_mgf}.

\textbf{(1) Graph encoders.}
Each modality $m\!\in\!\{\mathrm{EEG}, \mathrm{fMRI}\}$ is represented by a graph 
$(\tilde{\mathbf{A}}_m, \mathbf{X}_m)$, where $\tilde{\mathbf{A}}_m$ is the top-$k$ sparsified partial-correlation matrix and 
$\mathbf{X}_m = \mathbf{A}^{\mathrm{P}}_m + \psi(\mathbf{I})$ is the node feature matrix combining Pearson-based connectivity profiles with a learnable pseudo-identity prior.
Each graph is processed by a three-layer GraphConv encoder:
\begin{equation}
\mathbf{H}^{(l+1)}_m = \phi\!\left(\tilde{\mathbf{A}}_m \mathbf{H}^{(l)}_m \mathbf{W}^{(l)}_m\right)
\end{equation}
where $\phi$ is the ReLU activation, $\mathbf{H}^{(l)}_m$ the node features, and $\mathbf{W}^{(l)}_m$ the trainable weights at layer $l$ respectively.
Global mean pooling produces a subject-level embedding $\mathbf{z}_m \in \mathbb{R}^d$ for each modality.

\textbf{(2) Adaptive fusion gating.}
A two-layer MLP takes the concatenated embeddings $[\,\mathbf{z}_{\mathrm{EEG}} , \mathbf{z}_{\mathrm{fMRI}}\,]$ 
and outputs softmax weights $\alpha_{\mathrm{EEG}}$ and $\alpha_{\mathrm{fMRI}}$ satisfying 
$\alpha_{\mathrm{EEG}}+\alpha_{\mathrm{fMRI}}=1$.
The fused representation is computed as
\begin{equation}
\mathbf{z}_{\mathrm{fus}} = 
\alpha_{\mathrm{fMRI}} \, \mathbf{z}_{\mathrm{fMRI}} +
\alpha_{\mathrm{EEG}} \, \mathbf{z}_{\mathrm{EEG}}
\label{eq:adaptive_fusion}
\end{equation}
where gating weights are applied uniformly across feature dimensions after mean pooling.

\textbf{(3) Classification head.}
The fused embedding $\mathbf{z}_{\mathrm{fus}}$ is passed through a multilayer perceptron with sigmoid output:
\begin{equation}
\hat{y} = \sigma(\mathbf{W}_c \, \mathbf{z}_{\mathrm{fus}} + b_c)
\end{equation}
to predict the psilocybin versus no-psilocybin condition.
The model was implemented in \textbf{PyTorch} and optimised using
the binary cross-entropy (BCE) loss:

\vspace{-4mm}
\begin{equation}
\mathcal{L}_{\mathrm{BCE}} = -\frac{1}{N}\sum_{i=1}^{N}\!\left[y_i\log\hat{y}_i+(1-y_i)\log(1-\hat{y}_i)\right]
\end{equation}

\FloatBarrier
\section{Experimental Results}
\label{sec:results}

%\subsection{Modality Comparison Across Tasks}
% Table: Modality Comparison
\begin{table*}[!t]
\centering
\caption{EEG, fMRI, and Fusion performance on \textit{PsiConnect} (Meditation and Rest; 5-fold CV, BCE loss). 
Fusion is implemented using the proposed Brain--MGF network. Results are mean~$\pm$~SD (\%).}
\resizebox{\textwidth}{!}{
\begin{tabular}{l l c c c c c c c}
\toprule
\textbf{Task} & \textbf{Modality} & \textbf{Accuracy} & \textbf{Precision} & \textbf{Sensitivity} & \textbf{Specificity} & \textbf{F1 Score} & \textbf{ROC-AUC} & \textbf{BCE Loss} \\
\midrule
\multirow{3}{*}{Meditation} 
 & EEG & 59.64~$\pm$~9.45 & 65.18~$\pm$~13.03 & 55.38~$\pm$~16.74 & 64.39~$\pm$~19.63 & 58.26~$\pm$~12.02 & 67.80~$\pm$~10.35 & 0.623~$\pm$~0.038 \\
 & fMRI & 68.84~$\pm$~10.30 & 69.95~$\pm$~8.52 & 73.08~$\pm$~12.99 & 63.94~$\pm$~13.29 & 71.14~$\pm$~9.79 & 77.16~$\pm$~9.24 & 0.548~$\pm$~0.078 \\
 & \textbf{Fusion (EEG+fMRI)} & \textbf{74.02~$\pm$~6.99} & \textbf{74.05~$\pm$~9.00} & \textbf{80.90~$\pm$~12.93} & \textbf{66.21~$\pm$~17.22} & \textbf{76.48~$\pm$~6.95} & \textbf{75.72~$\pm$~11.29} & \textbf{0.580~$\pm$~0.108} \\
\midrule
\multirow{3}{*}{Rest} 
 & EEG & 68.53~$\pm$~7.91 & 71.76~$\pm$~7.29 & 62.95~$\pm$~15.29 & 73.94~$\pm$~6.29 & 66.57~$\pm$~11.52 & 75.35~$\pm$~6.23 & 0.566~$\pm$~0.065 \\
 & fMRI & 72.67~$\pm$~4.52 & 74.32~$\pm$~6.55 & 74.49~$\pm$~9.32 & 70.61~$\pm$~11.96 & 73.77~$\pm$~4.54 & 81.33~$\pm$~6.78 & 0.513~$\pm$~0.063 \\
 & \textbf{Fusion (EEG+fMRI)} & \textbf{76.00~$\pm$~8.15} & \textbf{76.30~$\pm$~10.30} & \textbf{79.10~$\pm$~11.11} & \textbf{72.58~$\pm$~12.41} & \textbf{77.20~$\pm$~8.20} & \textbf{85.83~$\pm$~8.57} & \textbf{0.467~$\pm$~0.115} \\
\bottomrule
\end{tabular}
}
\label{tab:results_psiconnect}
\end{table*}

We benchmarked \textbf{GCN} \cite{kipf2016semi}, \textbf{BrainGNN} \cite{li2021braingnn} and \textbf{BrainNetCNN} \cite{kawahara2017brainnetcnn} against three multimodal fusion variants:  

\begin{itemize}[leftmargin=12pt]
    \item \textbf{Late-Fusion (MLP):} Each modality is processed by a dedicated graph encoder to produce embeddings $\mathbf{z}_{\mathrm{fMRI}}$ and $\mathbf{z}_{\mathrm{EEG}}$. These are concatenated and passed to an MLP classifier:
    \begin{equation}
        \mathbf{z}_{\mathrm{fus}} = [\,\mathbf{z}_{\mathrm{fMRI}},\mathbf{z}_{\mathrm{EEG}}\,], \quad
        \hat{y} = \mathrm{MLP}(\mathbf{z}_{\mathrm{fus}})
    \end{equation}
    This late-fusion variant relies on dense layers to integrate EEG and fMRI information without explicit modality weighting.

    \item \textbf{Brain--MGF (w/o Gate):} A simplified version of the proposed model without (w/o) gating mechanism. The modality embeddings are directly averaged to form a unified representation:
    \begin{equation}
        \mathbf{z}_{\mathrm{fus}} = \tfrac{1}{2}\left(\mathbf{z}_{\mathrm{fMRI}} + \mathbf{z}_{\mathrm{EEG}}\right), \quad
        \hat{y} = \mathrm{MLP}(\mathbf{z}_{\mathrm{fus}}).
    \end{equation}
    This provides a fixed and interpretable fusion for isolating the effect of adaptive gating.

    \item \textbf{Brain--MGF (Proposed):} The full adaptive fusion framework, where a softmax gating network learns sample-specific weights $\alpha\!\in\!(0,1)$ to modulate modality contributions, as shown in Eq. \ref{eq:adaptive_fusion}.
    The gating enables context-sensitive integration of EEG and fMRI representations.
\end{itemize}

All models share identical preprocessing, 
%$k$-NN sparsification, 
top-40\% sparsification, per-modality ROI projections, and classifier depth for fair comparison. Training used the Adam optimiser (learning rate
$3\times10^{-4}$ to $5\times10^{-4}$,
weight decay $1\times10^{-4}$) with early stopping (patience 8 epochs). Evaluation employed \textbf{5-fold cross-validation} with stratified splits, reporting accuracy, precision, sensitivity, specificity, F1-score, ROC–AUC and BCE loss.

Table~\ref{tab:benchmark_meditation} compares Brain-MGF against established fMRI-based graph models on the meditation task. Among unimodal baseline models, GCN achieved the highest accuracy (69.71\%) and ROC-AUC (76.45\%), outperforming BrainGNN and BrainNetCNN, which exhibited higher loss variance and lower specificity.  
When extended to multimodal EEG--fMRI fusion, the proposed Brain-MGF framework consistently improved predictive performance. The full model achieved \textbf{74.02\%} accuracy and \textbf{76.48\%} F1-score, surpassing its Late-Fusion MLP and w/o Gate variants by
% comment: DLD again feels that this looks more like a subtraction than a range
4.3\% to 4.4\%.
This demonstrates the effectiveness of the adaptive gating mechanism in balancing complementary information across modalities while maintaining stable training loss 
% comment: DLD asks    0.58 of exactly what?
(0.58~$\pm$~0.11).

Table~\ref{tab:results_psiconnect} presents results for EEG, fMRI, and fusion models on both \textit{meditation} and \textit{rest} tasks.  
For meditation, fMRI yielded stronger discriminative performance (68.84\% accuracy) than EEG alone (59.64\%), consistent with the richer spatial information captured in fMRI connectivity.  
However, the multimodal fusion via Brain-MGF further enhanced accuracy to \textbf{74.02\%} and ROC-AUC to \textbf{75.72\%}, confirming that integrating temporal EEG dynamics with spatial fMRI structure benefits behavioural state decoding. For the rest condition, overall accuracies were higher across modalities, suggesting reduced inter-subject variability during unconstrained no-psilocybin scans. Fusion again achieved the best results (\textbf{76.00\%} accuracy, \textbf{85.83\%} ROC-AUC), highlighting the model’s robustness across cognitive states.  
In both tasks, the adaptive gating enabled effective cross-modality weighting, leading to consistent improvements over single-modality baseline models.

Across benchmarks and tasks, Brain-MGF achieved performance gains over unimodal and non-adaptive variants, providing preliminary evidence that adaptive EEG–fMRI fusion is useful.
%They validate Brain-MGF as a multimodal graph framework for decoding brain states under altered consciousness conditions such as psilocybin meditation.

\FloatBarrier
\vspace{-2mm}
\section{Discussion}
\label{sec:discussion}

%\subsection{Latent Embedding Visualisation}
% Figure: UMAP
\begin{figure*}[!t]
  \centering
  \begin{subfigure}{0.3\textwidth}
    \centering
    \includegraphics[width=\linewidth]{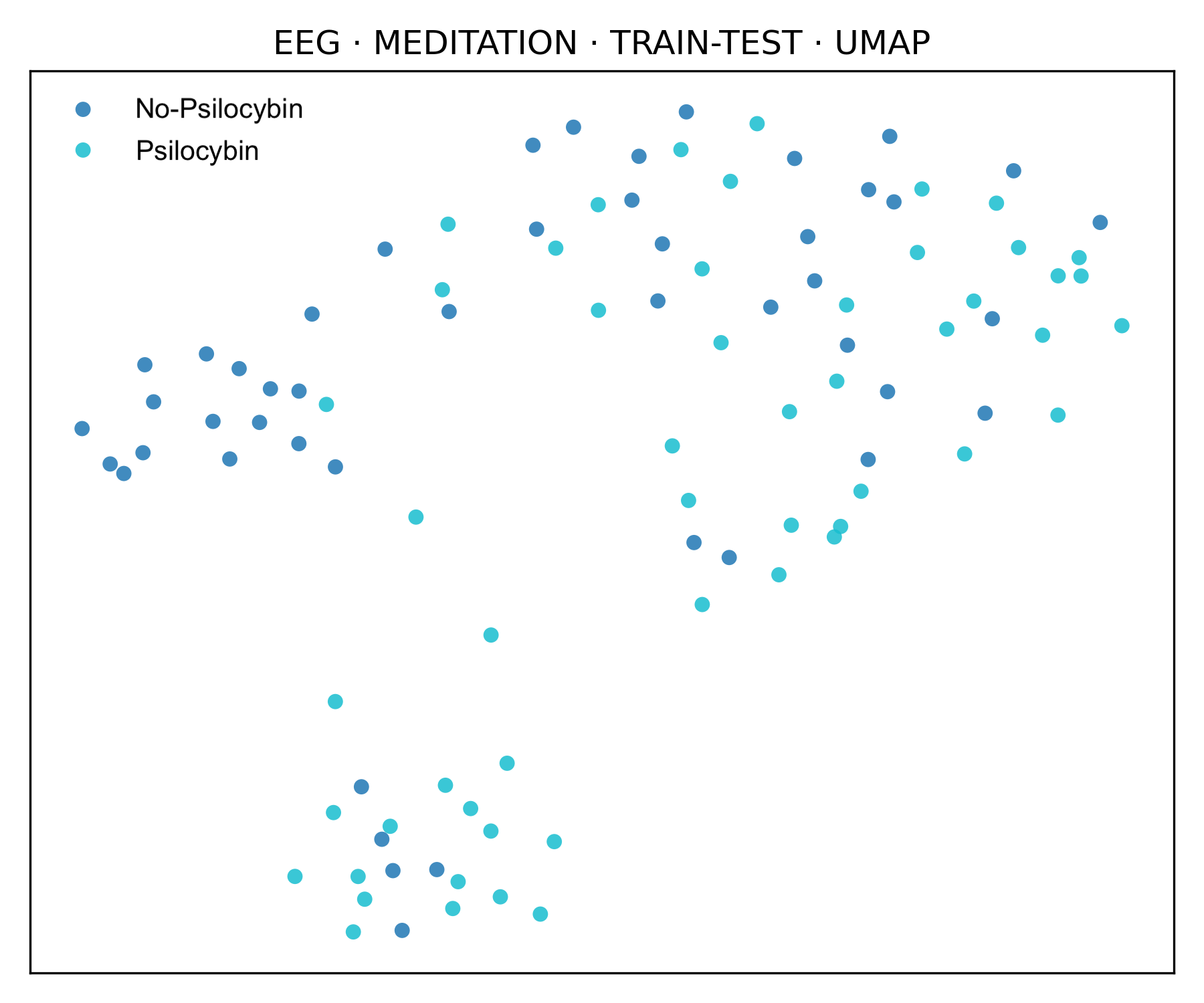}
    \caption{EEG}
    \label{fig:umap_eeg}
  \end{subfigure}
  \begin{subfigure}{0.3\textwidth}
    \centering
    \includegraphics[width=\linewidth]{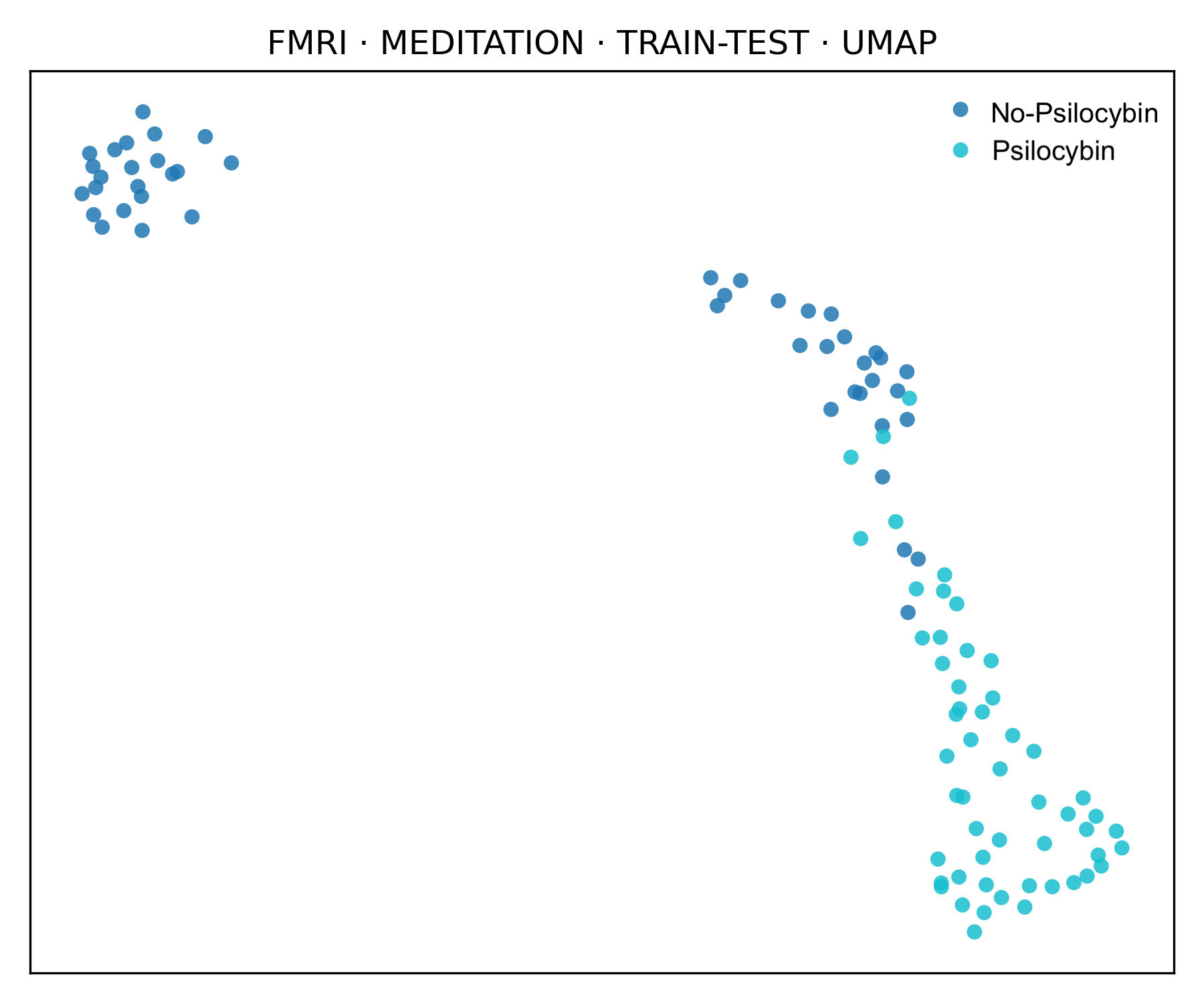}
    \caption{fMRI}
    \label{fig:umap_fmri}
  \end{subfigure}
  \begin{subfigure}{0.3\textwidth}
    \centering
    \includegraphics[width=\linewidth]{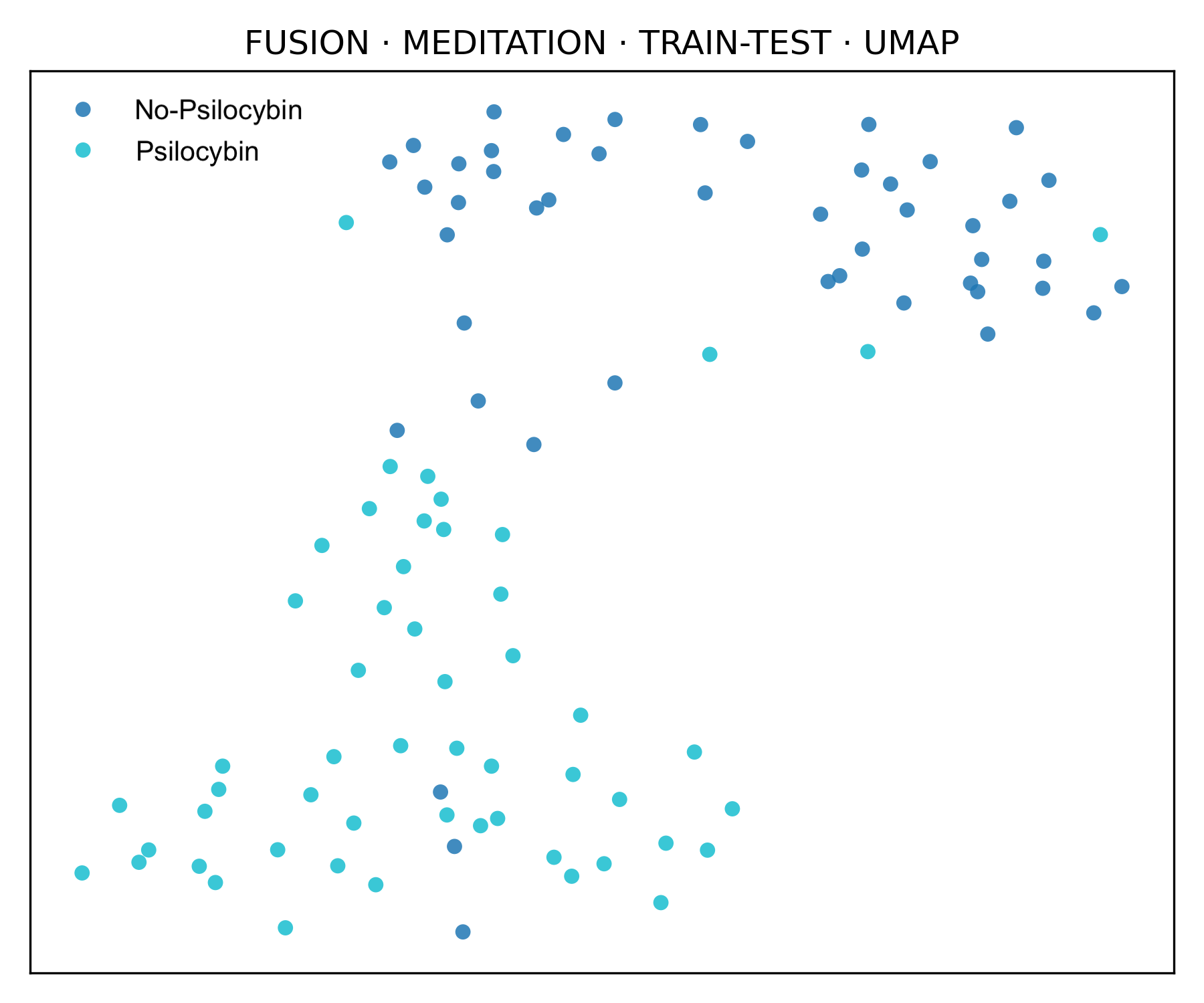}
    \caption{Fusion (EEG+fMRI)}
    \label{fig:umap_fusion}
  \end{subfigure}
  \caption{\textbf{UMAP visualisation of learned embeddings during meditation.}
  Each point represents a subject; colour denotes condition (no-psilocybin vs.\ psilocybin). 
  EEG embeddings show overlapping clusters, fMRI displays clearer separation, 
  and fusion achieves the most distinct manifolds, reflecting its quantitative advantage in Table~\ref{tab:benchmark_meditation} and Table~\ref{tab:results_psiconnect}.}
  \label{fig:umap_all}.
  \vspace{-5mm}
\end{figure*}

%The Brain-MGF network integrates EEG and fMRI connectivity to classify psilocybin brain states. The adaptive gating mechanism yielded modest but consistent improvements over unimodal and non-gated models, suggesting that feature-level fusion can better capture complementary temporal and spatial information. 

To qualitatively assess the latent representations, Fig.~\ref{fig:umap_all} visualises the final-layer embeddings of EEG, fMRI, and fused features for the meditation task using Uniform Manifold Approximation and Projection (UMAP) \cite{mcinnes2018umap}. UMAP is a non-linear dimensionality reduction technique that preserves both local and global data structure by constructing a fuzzy topological representation of the high-dimensional manifold. It was applied to the combined train–test embeddings ($n_{\mathrm{neighbors}}=15$,
% ${\mathrm{min_dist}}=0.1$,
${\mathrm{min_{dist}}}=0.1$,
Euclidean metric) from the best-performing fold for visualisation. While unimodal embeddings exhibit partial overlap between no-psilocybin and psilocybin, the fused embeddings form more distinct clusters, suggesting that adaptive fusion produces a more separable latent space.

%From a neurobiological perspective, psilocybin modulates large-scale integration by reducing default-mode network dominance and increasing cross-system coupling \cite{stoliker2025psychedelics}. Meditation further stabilises these network dynamics, yielding clearer separability of brain states. The fusion embeddings therefore capture context-dependent reorganisation consistent with recent evidence that psychedelics align neural activity with environmental context.

% One paragraph for quantifying the pychedelics changes?
Beyond aggregate performance, the adaptive gate provides an interpretable view of modality contributions. The gating coefficients $\alpha_{\mathrm{fMRI}}$ and $\alpha_{\mathrm{EEG}}$ were obtained from the softmax fusion defined in Eq.~\eqref{eq:adaptive_fusion}. Across folds, a modest fMRI bias was observed at rest (mean $\alpha_{\mathrm{fMRI}}=0.58\pm0.07$), with several folds significantly above chance (Wilcoxon vs.\ 0.5) \cite{cuzick1985wilcoxon}, suggesting that slow haemodynamic coupling offers a more stable representation of baseline integration. 
%The Wilcoxon signed-rank test is a non-parametric statistical method used to evaluate whether paired observations differ significantly from a reference value or median, without assuming normality.
During meditation, the average weighting still favoured fMRI (mean $\alpha_{\mathrm{fMRI}}=0.56\pm0.10$) but showed greater between-fold variability, including one EEG-dominant case, consistent with increased electrophysiological synchrony during focused attention. This state-dependent reweighting aligns with evidence that psilocybin reorganises large-scale network dynamics and enhances cross-system integration \cite{stoliker2025psychedelics}. Overall, the gating analysis indicates that fMRI connectivity forms a stable backbone for decoding psilocybin versus no-psilocybin, while EEG contributes complementary, context-dependent information.

% Overall, Brain-MGF proposes a multimodal graph framework for modelling altered-state connectivity, bridging fast EEG dynamics and slower fMRI topology. This approach could support future work linking neural graph structure to behavioural or phenomenological measures.
\vspace{-2mm}
\section{Conclusion}
\label{sec:conclusion}
This work presented \textbf{Brain–MGF}, a multimodal graph fusion framework for joint EEG–fMRI connectivity analysis under psilocybin. By learning adaptive weights between modalities, the model integrates complementary patterns of haemodynamic and electrophysiological connectivity in a data-driven manner. Experiments on the \textit{PsiConnect} dataset showed consistent improvements over unimodal and non-adaptive variants, indicating that adaptive gating enhances cross-modal representation of psilocybin-induced brain states. While preliminary, these findings suggest that Brain–MGF provides a practical approach for studying neural integration across physiological domains.

\section{Compliance with Ethical Standards}

The \textit{PsiConnect} study was approved by the Monash University Human Research Ethics Committee and registered under ACTRN12621001375842. All participants provided informed consent.

\bibliographystyle{IEEEbib}
\bibliography{strings,refs}

\end{document}